\documentclass[twocolumn,epsf,prb,amsmath,amssymb,showpacs,floatfix]{revtex4}
\usepackage{graphicx}
\usepackage{sidecap}
\begin{document}
\title{Electron-electron interactions in bilayer graphene quantum dots}
\author{M. Zarenia$^{1}$, B. Partoens$^{1}$, T. Chakraborty$^{2,1}$,
and F. M. Peeters$^{1}$}
\address{$^1$Department of Physics, University of Antwerp,
Groenenborgerlaan 171, B-2020 Antwerpen, Belgium.\\
$^2$Department of Physics and Astronomy, University of Manitoba,
Winnipeg, Canada R3T 2N2.}

\begin{abstract}
A parabolic quantum dot (QD) as realized by biasing nanostructured
gates on bilayer graphene is investigated in the presence of
electron-electron interaction. The energy spectrum and the phase
diagram reveal unexpected transitions as function of a magnetic
field. For example, in contrast to semiconductor QDs, we find a
novel valley transition rather than only the usual singlet-triplet
transition in the ground state of the interacting system. The origin
of these new features can be traced to the valley degree of freedom
in bilayer graphene. These transitions have important consequences
for cyclotron resonance experiments.
\end{abstract}

\pacs{81.05.ue, 73.21.La, 71.10.Li} \maketitle

\section{Introduction}
The electronic properties of quantum dots (QDs) in graphene, a single layer
of carbon atoms arranged in a honeycomb lattice \cite{geim,tapash_review,sarma}
have been studied extensively due to their unique properties and their potential
for applications in graphene devices \cite{ensslin,nori,hawrylak,matulis}. Since
Klein tunneling prevents electrostatic confinement in graphene, direct etching
of the graphene sheet is perhaps the only viable option for quantum confinement.
In such systems, controlling the shape and edges of the dot remains an important
challenge but the exact configuration of the edges is unknown \cite{ritter}. The
latter is important because the energy spectrum depends strongly on the type of
edges \cite{marko,zarenia1}.
\begin{figure}
\centering
\includegraphics[width=6 cm]{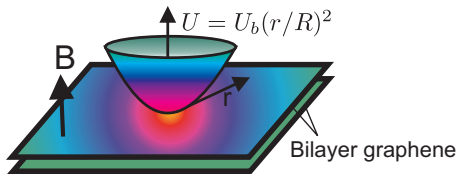}
\caption{(Color online) Schematic illustration of the potential profile for
a parabolic bilayer graphene quantum dot.}\label{fig1}
\end{figure}

Two coupled layers of graphene, called bilayer graphene (BLG), have
quite distinct properties from those of a single layer. In pristine
BLG the spectrum is gapless and is approximately parabolic at low
energies around the two nonequivalent points in the Brillouin zone
($K$ and $K'$). In a perpendicular electric field, the spectrum
displays a gap which can be tuned by varying the bias \cite{zhang}.
Nanostructuring the gate would allow tuning of the energy gap in BLG
which can be used to electrostatically confine QDs
\cite{milton1,milton2} and quantum rings \cite{zarenia}. Here the
electrons are displaced from the edge of the sample and consequently
edge disorder and the specific type of edges are no longer a
problem. Such gate defined QDs in BLG were recently fabricated by
different groups \cite{allen,QD2,muller}, who demonstrated
experimentally the confinement of electrons and Coulomb blockade.

In the present work we investigate the energy levels of a parabolic
QD in BLG in the presence of Coulomb interaction. Here we consider
the two-electron problem as the most simple case to investigate the
effect of electron-electron correlations. Similar studies have been
reported for semiconducting QDs over the last two decades
\cite{wagner,tapash} and recently for graphene QDs \cite{wunsch} and
graphene rings \cite{abergel}. At present no similar study has been
reported for BLG quantum dots. An important issue for graphene
structures is the extra valley-index degree of freedom where the
electrons have the possibility to be in the same valley or in
different valleys \cite{review,sabio}. Here we show that the
competition between the valley-index and the electron spin leads to
unique behaviors that shed light on the fundamental properties of
the ground state energy of BLG quantum dots.

\section{Continuum model}
In order to find the single-particle energy spectrum of a parabolic
QD we employ a four-band continuum model to describe the BLG sheet.
In the valley-isotropic form \cite{Falko}, the Hamiltonian is given
by
\begin{equation}\label{H}
H=\begin{pmatrix}
  \tau U(r) & \pi & t & 0 \\
 \pi^\dagger & \tau U(r )& 0 & 0\\
  t & 0 & -\tau U(r)& \pi^\dagger\\
 0 & 0 & \pi & -\tau U(r)
\end{pmatrix}
\end{equation}
where $t\approx 400$ meV is the inter-layer coupling term. The
additional coupling terms which lead to the trigonal warping effect
are neglected. The trigonal warping effect is only relevant at very
low energies (i.e. $E<2$  meV) in the absence of an electrostatic
potential \cite{Falko}. $\pi = -i\hbar v^{}_F
e^{i\theta}\big[\partial_r + i\partial_{\theta}/r-(eB/2
\hbar)r\big]$ is the momentum operator in polar coordinates and in
the presence of an external magnetic field $B$, and $v^{}_F = 10^6$
m/s is the Fermi velocity. The valley index parameter $\tau$
distinguishes the energy levels corresponding to the $K$ ($\tau=+1$)
and the $K'$ ($\tau=-1$) valleys. The electrostatic potential $U(r)$
is applied to the upper layer and $-U(r)$ to the lower layer. For
the QD profile we consider a parabolic potential
$U(r)=U^{}_b(r/R)^2$ where the potential $U^{}_b$ and the radius $R$
define the size of the dot [Fig.~\ref{fig1}]. The eigenstates of the
Hamiltonian (1) are the four-component spinors
\begin{equation}\label{psi1}
\psi(r,\theta)=e^{im\theta}\big[\phi_A(r),e^{-i\theta}\phi_B(r),
\phi_{B'}(r),e^{i\theta}\phi_{A'}(r) \big]^{T}
\end{equation}
where $\phi^{}_{A,B,B',A'}$ are the envelope functions associated
with the probability amplitudes at the respective sublattice sites
of the upper and lower graphene sheets and $m$ is the angular
momentum. The orbital angular momentum $L^{}_z$ does not commute
with the Hamiltonian and is no longer quantized. This is different
from two-dimensional semiconductor QDs, where $[H,L^{}_z]=0$.
However, the wave function $\psi(r,\phi)$ is an eigenstate of the
operator
\begin{equation}\label{Jz}
J^{}_z=L^{}_z+\left[\frac{\hbar}{2}\left(
                         \begin{array}{cc}
                           -\boldsymbol{I} & 0 \\
                           0 & \boldsymbol{I} \\
                         \end{array}
                       \right)
+\frac{\hbar}{2}\left(\begin{array}{cc}
                           \sigma_z & 0 \\
                           0 & -\sigma_z \\
                         \end{array}
                       \right)\right]
\end{equation}
with eigenvalue $m$, where $\boldsymbol{I}$  is the $2\times2$
unitary matrix and $\sigma^{}_z$ is one of the Pauli matrices. The
first operator inside the bracket is a layer index operator, which
is associated with the behavior of the system under inversion,
whereas the second one denotes the pseudospin components in each
layer.

Solving the Shr\"{o}dinger equation, $H\Psi=E\Psi$, the radial
dependence of the spinor components is described by
\begin{eqnarray}\label{eqS}
&&\Bigl[\frac{d}{d r} + \frac{m}{r} + \beta r\Bigr]\phi_A =
-\big[E-\tau U(r)\big]\phi_B,\cr &&\cr &&\Bigl[\frac{d}{d r} -
\frac{(m-1)}{r}- \beta r\Bigr]\phi_B = \big[E-\tau U(r)\big]\phi_A -
t\phi_{B'},\cr &&\cr && \Bigl[\frac{d}{d r} + \frac{(m+1)}{r}+ \beta
r\Bigr]\phi_{A'} = \big[E+\tau U(r)\big]\phi_{B'}-t\phi_A,\cr &&\cr
&& \Bigl[\frac{d}{d r} - \frac{m}{r}- \beta r \Bigr]\phi_{B'} =
-\big[E+\tau U(r)\big]\phi_{A'}
\end{eqnarray}
%
\noindent where $\beta=(eB/2\hbar)R^2$ is a dimensionless parameter.
The energy, the potential and the hopping term $t$ are written in
units of $E^{}_0=\hbar v^{}_F/R$ with $R$ being the unit of length.
The coupled equations (\ref{eqS}) are solved numerically using the
finite element method \cite{comsol}.
\begin{figure}
\centering
\includegraphics[width=9 cm]{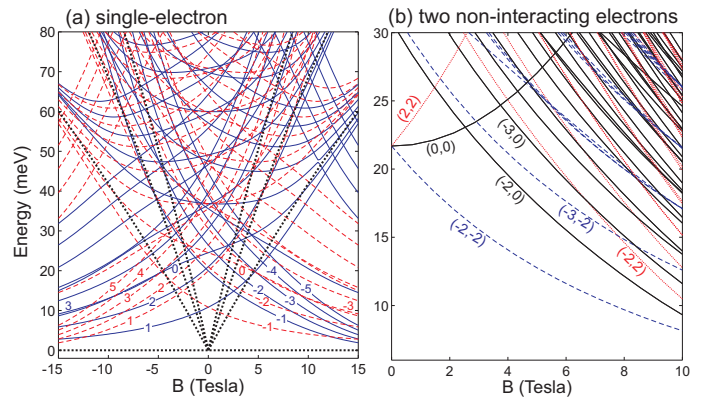}
\caption{(Color online) Energy spectrum of a parabolic QD in BLG
with $R=50$ nm and $U_b=150$ meV as a function of the magnetic field
for (a) single-particle and (b) two noninteracting electrons. The
lowest energy levels in (a) are labeled by the angular momentum $m$.
The levels corresponding to the $K$ and $K'$ valleys are
respectively shown by the blue solid and red dashed curves. The
black dotted curves are the Landau levels of a BLG sheet. The levels
in (b) are labeled by $(M,{\cal T})$, where $M=m_1+m_2$, is the
total angular momentum and ${\cal T}=\tau_1+\tau_2$ is the total
valley index. Levels having the same valley index are plotted using
the same type of curve.} \label{fig2}
\end{figure}

\subsection{Single-particle energy levels}
Figure \ref{fig2}(a) shows the lowest single-electron energy levels
as a function of the magnetic field for a QD with $U^{}_b=150$ meV
and $R=50$ nm. The energy levels are labeled by their angular
momentum and their valley index ($m,\tau$). We begin with the $B=0$
case. Notice that the single-particle ground state does not have
$m=0$ as expected for semiconductor QDs, but instead has the
momentum $m=1$ at $K$ and $m=-1$ at $K'$ in agreement with
Ref.~\cite{milton1}.

For large magnetic fields the eigenstates are strongly localized at
the origin of the dot, where $\Delta U\rightarrow0$. Therefore, the
spectrum approaches the Landau levels (LLs) of an unbiased BLG
(black dotted curves in Fig. \ref{fig2}(a)) and consequently some of
the energy levels approach $E=0$ as the field increases. Notice that
this is quite distinct from semiconductor QDs where the zeroth LL is
absent and thus the energy of the confined stats, i.e., the
Fock-Darwin states, increase with magnetic field. Breaking of the
electron-hole symmetry due to the presence of both electric and
magnetic fields lifts the valley-degeneracy in non-zero magnetic
fields. The energy spectrum also displays the
$E^{}_{K}(m,B)=E^{}_{K'}(-m,-B)$ symmetry, which is another feature
that is unique to BLG quantum dots. This symmetry is a consequence
of the fact that the QD is produced by a gate that introduces an
electric field and thus a preferential direction. Inserting
$(m,\beta)\rightarrow(-m, -\beta)$ and $\tau\rightarrow -\tau$ in
Eqs. (\ref{eqS}) and using $E^{}_{\tau}(m,B)=E^{}_{-\tau}(-m,-B)$,
one can find the relations $\phi_A^{K}(m,B)=\phi_{B'}^{K'}(-m,-B)$
and $\phi_{B}^{K}(m,B)=\phi_{A'}^{K'} (-m,-B)$ between the wave
function components of the $K$ and $K'$ valleys.
\begin{figure*}
\includegraphics[width=18 cm]{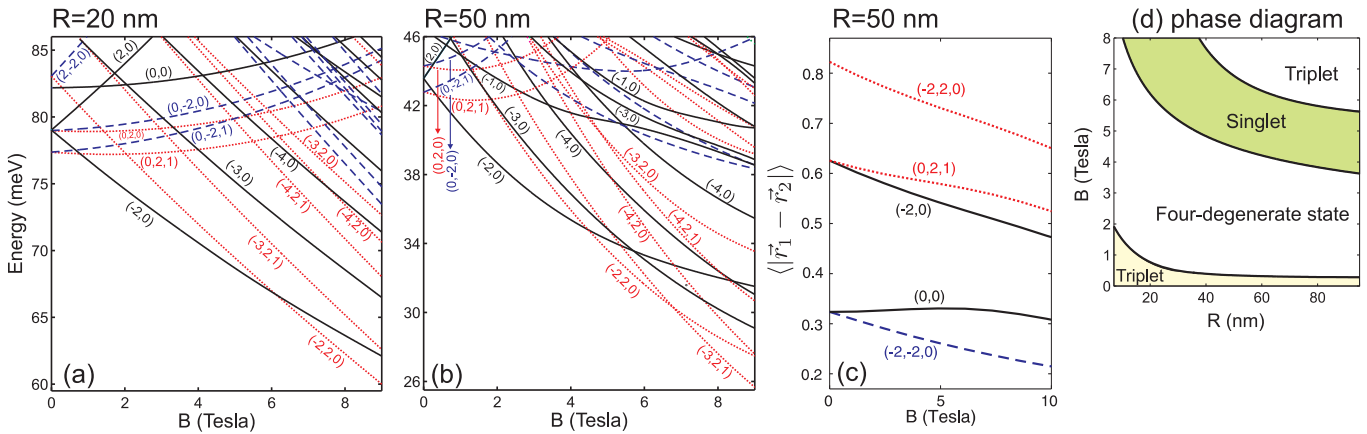}
\caption{(Color online) (a,b) The same as Fig.~\ref{fig2}(b) but in
the presence of Coulomb interaction and for (a) $R=20$ nm and (b)
$R=50$ nm. The levels are labeled by $(M,{\cal T},S)$ where $S$
indicates the total spin. The curves with the same total valley
index ${\cal T}$ are shown with the same type of curve. (c) The
average distance between the electrons as function of magnetic field
for different single particle basis functions. (d) The
radius-magnetic field ($R-B$) phase diagram of the ground state
energy of a two-electron BLG quantum dot.} \label{fig3}
\end{figure*}
\subsection{Two-electron energy spectrum} The Hamiltonian describing
the two-electron system is given by $H^{}_T=H({\bf r}^{}_1)+H({\bf
r}^{}_2)+V^{}_c({\bf r}^{}_1- {\bf r}^{}_2)$ where
$V^{}_c=e^2/(4\pi\kappa|{\bf r}^{}_1-{\bf r}^{}_2|)$ is the Coulomb
interaction between the two electrons with $\kappa$ being the
dielectric constant of BLG. In our calculations we set $\kappa=3.9$
which is the dielectric constant of gated BLG on top of a hexagonal
boron nitride (h-BN) substrate \cite{Dean}. We carry out an exact
diagonalization of the above Hamiltonian to obtain the eigenvalues
and eigenstates of the two-electron system. The corresponding
two-electron wave function with fixed total angular momentum $M$ and
total valley index ${\cal T}$ is constructed as linear combinations
of the one-electron wave functions:
\begin{equation}\label{psi12}
\Psi({\bf r}^{}_1,{\bf
r}^{}_2)=\sum_{i}^{N_s}\sum_{j}^{N_s}C_{ij}\Phi^{}_{i}({\bf
r}^{}_1)\otimes\Phi^{}_{j}({\bf r}^{}_2)
\end{equation}
where $\Phi$ is a eight-component wave function which is
$\Phi_{K}=[~\psi_{K},~0,~0,~0,~0~]^T$ corresponding to the $K$
valley and $\Phi_{K'}=[~0,~0,~0,~0,~\psi_{K'}~]^T$ corresponding to
the $K'$ valley \cite{bloch}. The four-component wave function
$\psi_{K(K')}$ is given by Eq.~(\ref{psi1}). Notice that the
two-electron wave function $\Psi({\bf r}^{}_1,{\bf r}^{}_2)$ has
sixty-four components. The subscripts $i\equiv(m^{}_i,\tau^{}_i)$
and $j\equiv(m^{}_j,\tau^{}_j)$ correspond to the one-electron
energy levels where the summations in Eq.~(\ref{psi12}) are such
that the relations $M=m^{}_i+m^{}_j$ and ${\cal
T}=\tau^{}_i+\tau^{}_j$ are satisfied. In our calculations $N^{}_s$,
i.e. the number of lowest single-particle states, is chosen
sufficiently large to guarantee the convergence of the energies. The
singularity due to the $1/|{\bf r}^{}_1- {\bf r}^{}_2|$ term in the
matrix elements is avoided by using an alternative expression in
terms of the Legendre function of the second kind of half-integer
degree \cite{cohl}.

In Fig.~\ref{fig3}, we show two representative spectra for two
interacting electrons in a BLG quantum dot with radius (a) $R=20$
nm, and (b) $R=50$ nm, and $U^{}_b=150$ meV. To clearly see the
effect of electron correlations, the spectra for two non-interacting
electrons in a dot with $R=50$ nm is shown in Fig.~\ref{fig2}(b) for
comparison. The levels are labeled by $(M,{\cal T},S)$ with
$M=m^{}_1+m^{}_2$ the total angular momentum, ${\cal T}=\tau^{}_1
+\tau^{}_2$ the total valley index, and $S$ the total spin. Energy
levels with the same $\cal T$ are drawn using the same type of
curve. Two electrons can form a non-degenerate singlet state ($S=0$)
and a three-fold degenerate triplet state ($S=1$). In case the
quantum number $S$ is omitted, the singlet and triplet states are
degenerate. 
In the following discussion it is useful to characterize the
many-body state by the single particle basis function in expression
(\ref{psi12}) which has the largest contribution. We denote the
basis function in which the first and second electron have,
respectively, angular momentum $m_1$ and $m_2$ and valley index
$\tau_1$ and $\tau_2$ as $(m_1,\tau_1) \otimes (m_2,\tau_2) \equiv
\phi_{m_1,\tau_1}\otimes \phi_{m_2,\tau_2}$.

The spectra of the two interacting electrons in Figs. \ref{fig3}(a)
and \ref{fig3}(b) are a result of three competing effects. It is
evident that the energy of the single particle states as function of
the magnetic field, shown in Fig. \ref{fig2}(b), determines partly
the spectrum. However, turning on the Coulomb interaction between
the electrons changes the spectrum drastically. While the
non-interacting state $(-2,-2) \equiv (-1,-1) \otimes (-1,-1)$ is
the ground state, the many-body interacting state in which this
single particle basis function has the largest contribution is an
excited state and does not even appear in Figs. \ref{fig3}(a) and
\ref{fig3}(b). Instead the many-body state $(0,2,1)$ is the ground
state for small magnetic field values, with the main contribution
$(1,1) \otimes (-1,1)$. The Coulomb interaction is clearly not a
small perturbation. In Fig. \ref{fig3}(c), the evolution of the
average distance between both electrons is shown for different
single particle basis states (for the $R=50$ nm case). This
difference in the average distance can be understood from the single
particle densities. While single particle states $(-1,-1)$ and
$(1,1)$ have a non-zero density in the origin, the density of the
single particle state $(-1,1)$ is zero in the origin. Therefore,
this average distance is much larger, and consequently the Coulomb
interaction is much lower, for the basis function $(1,1)\otimes
(-1,1)$ than for $(-1,-1)\otimes (-1,-1)$, which is the reason why
the many-body state $(0,2,1)$ has a lower energy.

A more subtle effect is played by the exchange interaction. As
mentioned, the ground state at small magnetic field values is given
by the many-body state $(0,2,1)$, which is a triplet state. The
corresponding singlet state $(0,2,0)$ is slightly higher in energy.
Also note that the state $(-2,0)$, with the main single particle
contribution $(-1,1)\otimes(-1,-1)$ is higher in energy at small
magnetic fields, although the Coulomb interaction contribution is
expected to be very similar (compare curves $(0,2,1)$ and $(-2,0)$
in Fig. \ref{fig3}(c)). The reason is again the exchange interaction
energy gain for the triplet state $(0,2,1)$. State $(-2,0)$ is
fourfold degenerate: the triplet configuration does not lead to an
exchange energy gain because both electrons occupy different
valleys. Only for larger magnetic field values, the state $(-2,0)$
takes over from the state $(0,2,1)$ to become the ground state. This
is caused by the evolution of the single particle energies. The next
state that becomes the ground state with increasing magnetic field
is the singlet $(-2,2,0)$, with main single particle contribution
$(-1,1)\otimes(-1,1)$. Because twice the same single particle level
is occupied, no exchange energy gain is possible. Nevertheless, this
state becomes the ground state due to the evolution of the single
particle energies, together with the fact that the average distance
between both electrons is even smaller (see Fig. \ref{fig3}(c)), as
both electrons occupy a single particle level with zero density in
the origin. With increasing field, it becomes beneficial for an
electron to jump from the single particle level $(-1,1)$ to the
single particle level $(-2,1)$, resulting in the triplet state
$(-3,2,1)$ with main contribution $(-2,1)\otimes(-1,1)$.

While in conventional semiconductor QDs, the ground state shows a
series of singlet to triplet transitions as function of the magnetic
field strength \cite{tapash,wagner}, a more complex phase diagram is
found for BLG QDs. In Fig. \ref{fig3}(d) we plot this phase diagram
for the same $U_b$ and dielectric constant as used for Fig.
\ref{fig3}(a,b). For small magnetic field values, the ground state
is found to be a triplet state. With increasing field a valley
transition occurs, resulting in a fourfold-degenerate state. Next,
again a valley transition occurs into a singlet state. Further
increasing the magnetic field favors again a triplet state.
\begin{figure}
\centering
\includegraphics[width=8.5 cm]{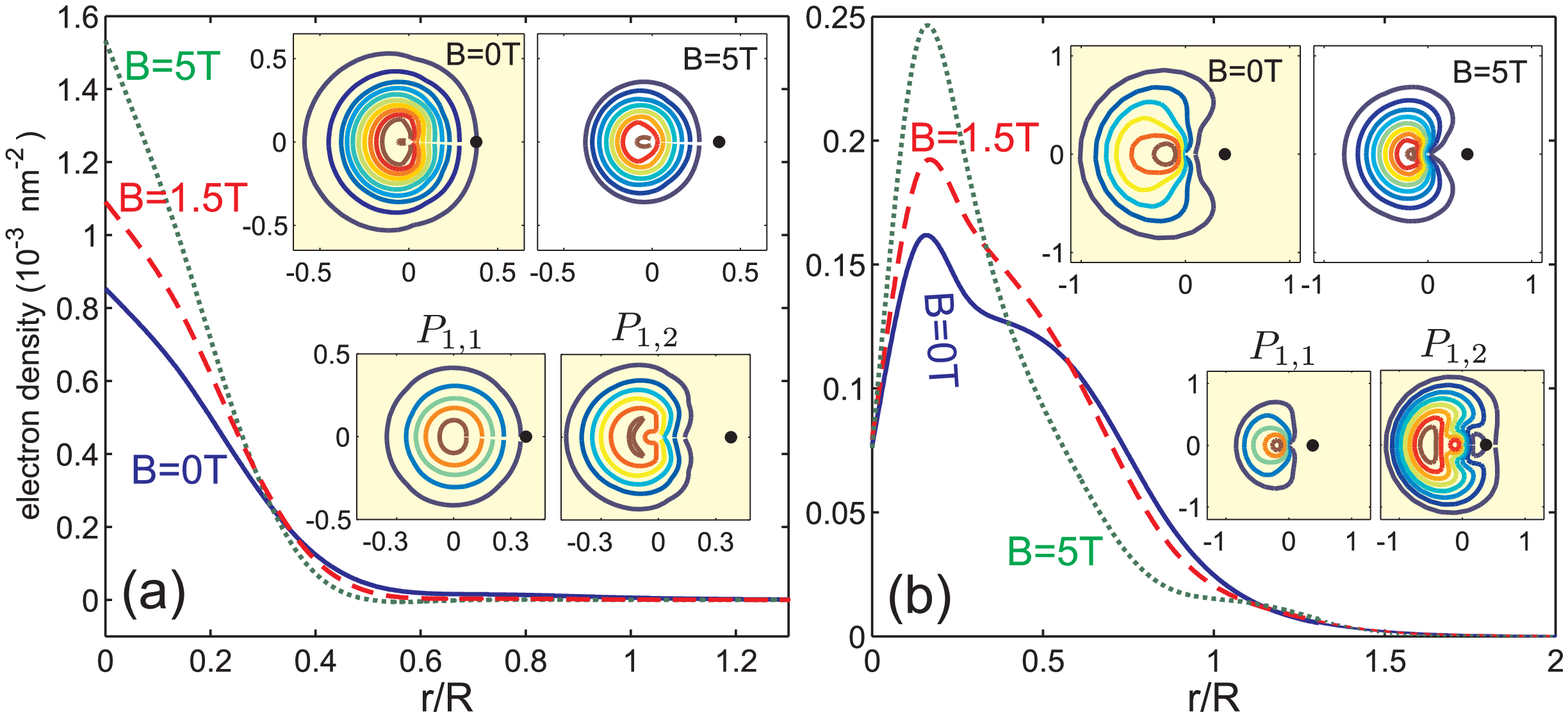}
\caption{(Color online) The radial electron density for the
two-electron QD of Fig.~\ref{fig3}(b) for (a) $(M,{\cal
T})\equiv(-2,0)$ and (b) $(M,{\cal T},S)\equiv(-3,2,1)$ and for
$B=0$, $1.5$, $5$ T. The upper insets show the total
pair-correlation function for $B=0$ T and $B=5$ T. The lower insets
show separate parts of the pair-correlation function for $B=0$. The
black dot indicates the position of the first electron which is
pinned at ${\bf r}^{}_1=(0.4R,0)$.} \label{fig4}
\end{figure}

The electron density, $\rho(r)=\sum_{i=1}^{2}\langle\delta( {\bf
r}-{\bf r}^{}_i)\rangle$ is shown in Figs.~\ref{fig4}(a) and
\ref{fig4}(b), respectively for the $(-2,0)$ and $(-3,2,1)$ states,
of a two-electron BLG quantum dot with $R=50$ nm, $U^{}_b=150$ meV,
and for three values of the external magnetic field $B=0,1.5,5$ T.
Comparing the density profiles in (a) and (b), the maximum of the
density is shifted towards higher radial distance in the state
$(-3,2,1)$. This is a consequence of the fact that the electrons in
the many-particle $(-3,2,1)$ state occupy single-particle states
with higher angular momentum. As the magnetic field increases, the
electrons are pulled closer towards the center of the dot. The upper
insets in Figs.~\ref{fig4}(a) and \ref{fig4}(b) show the total
pair-correlation function $P=|\Psi({\bf r}^{},{\bf
r}^{}_2)|^2=P_{1,1}+P_{1,2}$, for $B=0$ T and $B=5$ T. The
$P_{1,1}$, and $P_{1,2}$ terms refer respectively to the
contribution of the sublattices in the same layer and in the
different layers. In the lower insets the terms $P_{1,1}$ and
$P_{1,2}$ are plotted separately for $B=0$ T which respectively
indicate the probability to find the second electron in the same
layer, or in the other layer if the first electron is fixed at a
certain point.
\begin{figure}
\centering
\includegraphics[width=8cm]{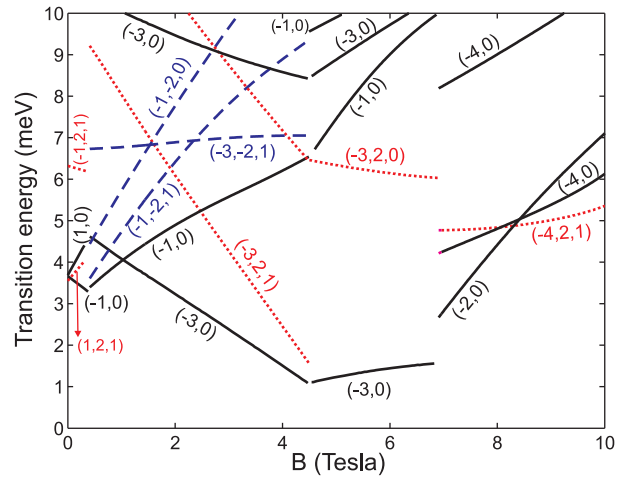}
\caption{(Color online) The cyclotron transition energies for the QD
of Fig.~\ref{fig3}(b). The transition energies are labeled by their
final states: $(M,{\cal T},S)$ for the triplet (or singlet) states
and $(M,{\cal T})$ for the degenerate single-triplet states. Levels
having the same total valley index are plotted using the same type
of curve.} \label{fig5}
\end{figure}

In cyclotron resonance experiments, transitions are induced between
the ground state and excited states. For the BLG quantum dot the
selection rule on the change of total momentum $\Delta m=\pm1$ is
still fulfilled. This is apparent when we calculate the transition
energies and the corresponding transition rates for dipole
transitions using the relation $|\langle\Psi^{}_i|
\sum_{j=1}^{2}r^{}_j \exp{(\pm
i\theta^{}_j)}/2|\Psi^{}_f\rangle|^2$. This relation also implies
conservation of the total spin, i.e., $\Delta S=0$. The valley
degree of freedom dictates new transition rules for BLG quantum
dots, i.e., $\Delta{\cal T}=0$ or $\Delta{\cal T}=\pm2$. This means
that those transitions are possible in which at least one electron
remains in the same valley during the transition. The lowest
possible transition energies for a two-electron quantum dot with
$R=50$ nm and $U_b=150$ meV are shown in Fig.~\ref{fig5}. The
possible transitions are labeled by the final states ($M,{\cal
T},S$). The discontinuities between the transition energies at
$B\approx 0.5$ T, $B\approx 0.5$ T and $B\approx 7$ T are due to the
valley and singlet-triplet transitions (see Fig. \ref{fig3}(b)).
\section{CONCLUDING REMARKS}
In summary, we have investigated the energy levels, the electron
density, the pair correlation function and the cyclotron transition
energies of electrostatically confined QDs containing one or two
electrons in a BLG. Such QDs can be realized experimentally by using
nanostructured gate potentials on a BLG. In contrast to conventional
semiconductor QDs, we found that the ground state energy of the
two-electron spectrum exhibits a valley transition rather than a
spin singlet-triplet transition. This is due to the extra valley
degree of freedom in BLG in which the electrons can be in different
valleys and thereby allowing the four-degenerate single-triplet
states as the ground state. Experimental confirmation of our
prediction can come from spin susceptibility measurements.
\section*{ACKNOWLEDGMENTS}
This work was supported by the Flemish Science Foundation (FWO-Vl),
the European Science Foundation (ESF) under the EUROCORES program
EuroGRAPHENE (project CONGRAN), and the Methusalem foundation of the
Flemish Government. T.C. is supported by the Canada Research Chairs
program of the Government of Canada.


\begin{thebibliography}{9}
\bibitem{bloch}
The total single particle wave functions can be expressed as a
superposition of the contribution of the two valleys via the Bloch
wave function, i.e.
$\exp({i\boldsymbol{K}\cdot\boldsymbol{r}})\Phi_K$ and
$\exp({i\boldsymbol{K'}\cdot\boldsymbol{r}})\Phi_{K'}$. When both
electrons are in the same valley the Bloch terms cancel each other
and when the electrons belong to different valleys the corresponding
matrix element becomes zero
$\langle\Phi^{K}(\boldsymbol{r_1})\otimes\Phi^{K'}_j(\boldsymbol{r_2})|V_c|\Phi^{K'}
(\boldsymbol{r_1})\otimes\Phi^{K}(\boldsymbol{r_2})\rangle=0$. Thus
the Bloch wave functions are not important when constructing the
matrix elements.
\end{thebibliography}
\end{document}